\documentclass[aps,amssymb,amsmath,pra,twocolumn,showpacs,superscriptaddress,longbibliography]{revtex4-1}

\usepackage{graphicx}
\usepackage{dcolumn}
\usepackage{bm}
\usepackage{subfigure}
\usepackage{color}

\usepackage{amssymb,amsmath,amsfonts,latexsym,graphicx,verbatim}

\usepackage[margin=0.6in]{geometry}

\usepackage[english]{babel}
\usepackage{times}
\usepackage{dsfont}
\usepackage{latexsym}
\usepackage{fancyhdr}
\usepackage{float}
\usepackage{afterpage}
\usepackage{enumitem}
\usepackage{mleftright}

\usepackage{eso-pic, graphicx}

\usepackage{listings}
\usepackage{multirow}
\usepackage{xcolor,colortbl}

\usepackage{bbm}
\usepackage{upgreek}
\usepackage{amsmath}
\usepackage{newtxtext,newtxmath}

\definecolor{Ablue}{rgb}{0.96,0.24,0.00}

\definecolor{Abluetitle}{rgb}{0.,0.24,0.51}
\newcommand{\bluetitle}{\color{Abluetitle}}

\definecolor{orange}{rgb}{0.96,0.24,0.00}

\definecolor{darkred}{rgb}{0.55, 0.0, 0.0}

\definecolor{darksalmon}{rgb}{0.91, 0.59, 0.48}
\definecolor{maroon}{cmyk}{0,0.87,0.68,0.32}

\definecolor{mustard}{rgb}{1.0, 0.86, 0.35}

\newcommand{\mc}[2]{\multicolumn{#1}{|c|}{#2}}
\definecolor{Gray}{gray}{0.85}
\definecolor{LightCyan}{rgb}{0.88,1,1}
\newcolumntype{a}{$>${\columncolor{Gray}}c}
\newcolumntype{b}{$>${\columncolor{white}}c}

\usepackage{array}
\newcolumntype{L}[1]{$>${\raggedright\let\newline\\\arraybackslash\hspace{0pt}}m{#1}}
\newcolumntype{C}[1]{$>${\centering\let\newline\\\arraybackslash\hspace{0pt}}m{#1}}
\newcolumntype{R}[1]{$>${\raggedleft\let\newline\\\arraybackslash\hspace{0pt}}m{#1}}

\usepackage[colorlinks=true , citecolor=blue,urlcolor=blue]{hyperref}

\newcommand{\us}{$\mu$s}

\newcommand{\id}{\mathds{1}}

\newcommand{\NV}{NV$^{\text{-}}$\:}

\newcommand{\xd}{\delta}

\newcommand{\vxe}{\varepsilon}

\newcommand{\CC}{\R{CC}}

\newcommand{\xtopt}{\xt_{\R{opt}}}
\newcommand{\tacq}{t_{\R{acq}}}
\newcommand{\xg}{\gamma}
\newcommand{\xt}{\theta}

\newcommand{\xn}{\eta}

\newcommand{\xr}{\rho}
\newcommand{\xo}{\omega}

\newcommand{\app}{\approx}

\newcommand{\Bp}{B_{\R{pol}}}

\newcommand{\Cs}{{}^{13}\R{C}}

\newcommand{\mG}[0]{\mathcal{G}}

\newcommand{\xD}{\Delta}

\newcommand{\xO}{\Omega}

\newcommand{\fr}[2]{\frac{#1}{#2}}
\newcommand{\ov}[1]{\overline{#1}}
\newcommand{\wt}[1]{\widetilde{#1}}

\newcommand{\s}[2]{\sum_{#1}^{#2}}

\newcommand{\mH}[0]{\mathcal{H}}

\newcommand{\mS}[0]{\mathcal{S}}

\newcommand{\rt}{\rightarrow}

\newcommand{\dg}{\dagger}

\newcommand{\beq}{\begin{equation}}
\newcommand{\eeq}{\end{equation}}
                  
\newcommand{\benum}{\begin{enumerate}}
\newcommand{\eenum}{\end{enumerate}}
                    
\newcommand{\bit}{\begin{itemize}}
\newcommand{\eit}{\end{itemize}}

\newcommand{\bea}{\begin{eqnarray}}
\newcommand{\eea}{\end{eqnarray}}

\newcommand{\bev}{\begin{verbatim}}
\newcommand{\eev}{\end{verbatim}}

\newcommand{\non}{\nonumber}

\newcommand{\zt}{\times}

\newcommand{\qt}{\tau}

\newcommand{\lb}{\left(}
\newcommand{\rb}{\right)}
\newcommand{\lsb}{\left[}
\newcommand{\rsb}{\right]}

\newcommand{\labs}{\left|}
\newcommand{\rabs}{\right|}

\newcommand{\T}[1]{\textbf{#1}}
\newcommand{\I}[1]{\textit{#1}}
\newcommand{\R}[1]{\textrm{#1}}

\newcommand{\zl}[1]{\label{eqn:#1}}
\newcommand{\zr}[1]{Eq. (\ref{eqn:#1})}
\newcommand{\zfl}[1]{\protect\label{fig:#1}}
\newcommand{\zfr}[1]{Fig. \ref{fig:#1}}
\newcommand{\ztl}[1]{\label{table:#1}}
\newcommand{\ztr}[1]{Table \ref{table:#1}}

\newcommand{\expec}[1]{\left\langle #1\right\rangle}

\newcommand{\ba}{\left\{ \begin{array}{lr}}
\newcommand{\ea}{\end{array}\right.}

\newcommand{\BRd}[1]{\textcolor{red}{#1}} 

\newcommand{\Tr}[1]{\textrm{Tr}\left\{{#1}\right\}}

\newcommand{\blist}[1]{
 \begin{list}{#1}
 \begin{align}
	 arrow
 \end{align}
 $\checkmark\star
  { \setlength{\itemsep}{3pt}
     \setlength{\parsep}{2pt}
     \setlength{\topsep}{3pt}
     \setlength{\partopsep}{0pt}
     \setlength{\leftmargin}{1em}
     \setlength{\labelwidth}{1em}
     \setlength{\labelsep}{0.5em} } }
\newcommand{\elist}{
  \end{list}  }

\DeclareMathSymbol{\vartheta}{\mathalpha}{letters}{"12}
\DeclareMathSymbol{\theta}{\mathalpha}{letters}{"23}
\DeclareMathSymbol{\phi}{\mathalpha}{letters}{"27}
\DeclareMathSymbol{\varphi}{\mathalpha}{letters}{"1E}

\newcommand{\bef}
{
\begin{figure}[htbp]
\centering
}

\newcommand{\eef}{\end{figure}}

\mleftright
\def\approxprop{%
  \def\p{%
    \setbox0=\vbox{\hbox{$\propto$}}%
    \ht0=0.6ex \box0 }%
  \def\s{%
    \vbox{\hbox{$\sim$}}%
  }%
  \mathrel{\raisebox{0.7ex}{%
      \mbox{$\underset{\s}{\p}$}%
    }}%
}

\newcommand{\beginsupplement}{%
        \setcounter{table}{0}
        \renewcommand{\thetable}{S\arabic{table}}%
        \setcounter{figure}{0}
        \renewcommand{\thefigure}{S\arabic{figure}}%
				
     }

\newcommand{\affA}{Department of Chemistry, University of California, Berkeley, Berkeley, CA 94720, USA.}
\newcommand{\affB}{Centro Brasileiro de Pesquisas Físicas, Rua Dr. Xavier Sigaud 150, 22290-180 Rio de Janeiro, Rio de Janeiro, Brazil.}
\newcommand{\affC}{Department of Physics and CUNY-Graduate Center, CUNY-City College of New York, New York, NY 10031, USA.}
\newcommand{\affD}{Energy Geoscience Division, Lawrence Berkeley National Laboratory, Berkeley, CA 94720, USA.}
\newcommand{\affE}{Fakult\"{a}t Physik, Technische Universit\"{a}t Dortmund, D-44221 Dortmund, Germany.}
\newcommand{\affF}{Department of Chemical and Biomolecular Engineering, and Materials Science Division Lawrence Berkeley National Laboratory, University of California, Berkeley, Berkeley, CA 94720, USA.}

\begin{document}
\title{Dynamical decoupling in interacting systems: applications to signal-enhanced hyperpolarized readout}

	
	\author{A. Ajoy}\email{ashokaj@berkeley.edu}\affiliation{\affA}
	\author{R. Nirodi}\affiliation{\affA}
	\author{A. Sarkar}\affiliation{\affA}
	\author{P. Reshetikhin}\affiliation{\affA}
	\author{E. Druga}\affiliation{\affA}
	\author{A. Akkiraju}\affiliation{\affA}
	\author{M. McAllister}\affiliation{\affA}
	\author{G. Maineri}\affiliation{\affA}
\author{S. Le}\affiliation{\affA}
\author{A. Lin}\affiliation{\affA}
\author{A. M. Souza}\affiliation{\affB}	
\author{C. A. Meriles}\affiliation{\affC}
  \author{B. Gilbert}\affiliation{\affD}
	 \author{D. Suter}\affiliation{\affE}
	\author{J. A. Reimer}\affiliation{\affF}
	\author{A. Pines}\affiliation{\affA}

\begin{abstract}
Methods that preserve coherence broadly impact all quantum information processing and metrology applications. Dynamical decoupling methods accomplish this by protecting qubits in noisy environments but are typically constrained to the limit where the qubits themselves are non-interacting. Here we consider the alternate regime wherein the inter-qubit couplings are of the same order as dephasing interactions with the environment. We propose and demonstrate a multi-pulse protocol that protects transverse spin states by suitably Hamiltonian engineering the inter-spin coupling while simultaneously suppressing dephasing noise on the qubits. We benchmark the method on $\Cs$ nuclear spin qubits in diamond, dipolar coupled to each other and embedded in a noisy electronic spin bath, and hyperpolarized via optically pumped \NV centers. We observe effective state lifetimes of $\Cs$ nuclei $T_2'\app$2.5s at room temperature, an extension of over 4700-fold over the conventional $T_2^{\ast}$ free induction decay. The spins are continuously interrogated during the applied quantum control, resulting in $\Cs$ NMR line narrowing and an $>$500-fold boost in SNR due to the lifetime extension. Together with hyperpolarization spin interrogation is accelerated by $>10^{11}$ over conventional 7T NMR. This work suggests strategies for the dynamical decoupling of coupled qubit systems with applications in a variety of experimental platforms.
\end{abstract}

\maketitle

\T{\I{Introduction}} --  Protecting coherence in quantum systems is an undertaking of immense importance in all quantum technologies, especially related to metrology and simulation~\cite{Nielsen00b}. Control techniques that coherently manipulate quantum systems have proved valuable as a means to suppress qubit noise, and can result in extensions of native $T_2^{\ast}$ qubit coherence decay times to far longer \I{engineered} $T_2$ coherence times~\cite{Suter16}. Indeed, $\left|T_2/T_2^{\ast}\right|$ ratios are now routinely over two orders of magnitude in a variety of physical quantum architectures~\cite{Du09,Ryan10,Bylander11,Naydenov11,Biercuk09,Medford12,Pokharel18}. Dynamical decoupling (DD)~\cite{Viola99b,Viola98,lidar98,Khodjasteh05} has come to typify the most widely employed qubit coherence and state preservation techniques. These methods consider individual (\I{non-interacting}) qubits coupled with an environment (\I{bath}), and prescribe means to protect them against decoherence by manipulating qubit degrees of freedom alone~\cite{Witzel07}. Extensions have allowed precise spectroscopic characterization of the noisy environment the qubits are embedded in~\cite{Alvarez11,Bylander11,Sung19}, and have engendered new techniques for quantum sensing~\cite{Degen17}.

\begin{figure}[t]
  \centering
  {\includegraphics[width=0.45\textwidth]{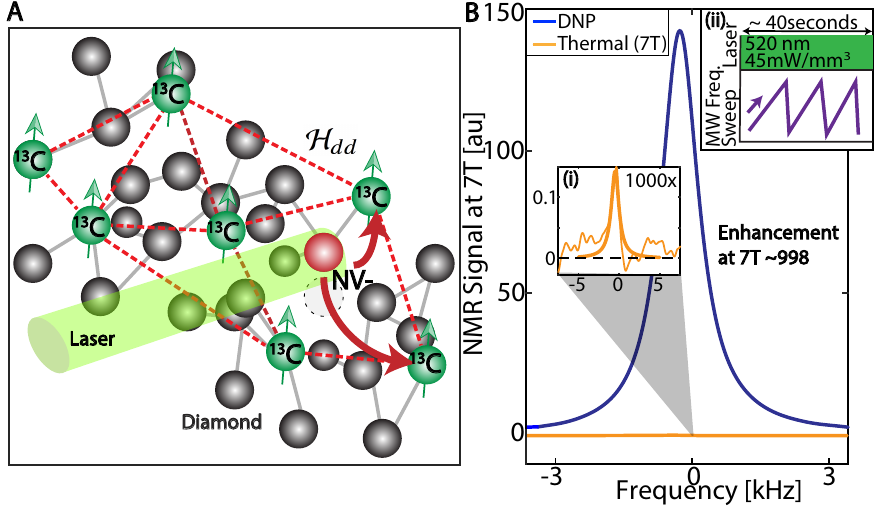}}
  \caption{\T{System and hyperpolarization.} (A) Dipolar coupled $\Cs$ nuclear spins in diamond doped with NV centers. Dashed lines denote couplings, and red arrows illustrate NV-mediated polarization transfer transfer that yields highly polarized $\Cs$ nuclei~\cite{Ajoy18}.  (B) \I{Representative DNP results} in a natural abundance $\Cs$ crystal illustrates the FID signal enhancement ($\vxe\app$998 over 7T). Here blue (orange) lines refer to the hyperpolarized (7T thermal) signals. (i) \I{Inset}: zoomed version of the flat (orange) line in the main panel, showing conventional 7T NMR signal. (ii) \I{Hyperpolarization protocol}: chirped MW irradiation is applied under optical pumping at $\Bp\sim$38mT. Sample is subsequently shuttled to $B_0=$7T where the $\Cs$ nuclei are interrogated.}
\zfl{fig1}
\end{figure}

The alternate regime wherein the collection of qubits are non-negligibly interacting is important to applications where multi-qubit systems are harnessed for metrology~\cite{Goldstein11} or simulation~\cite{Preskill18} yet are outside the ambit of conventional DD. The limit where the inter-qubit interaction $\mH_I$ is comparable to, or exceeds, dephasing interactions due to the bath $\mH_z$ is particularly pertinent. In this paper, we propose a simple dynamical decoupling protocol in this regime, that acts to preserve qubit states along a predetermined transverse Bloch-sphere axis.
We refer to the corresponding decay time as $T_2'$ to distinguish from \I{coherence} time $T_2$ for preservation of arbitrary qubit states.
In prototypical dipolar-coupled  $\Cs$ nuclear spins in diamond at room-temperature (see \zfr{fig1}A), we demonstrate lifetime extension by three orders of magnitude from $T_2^{\ast}=$517$\mu$s to $T_2'=$2.45s.  The 75MHz qubits, in this case, perform $>10^8$ precessions in a magnetic field before decaying. The $\Cs$ spins are continuously interrogated during the applied quantum control, allowing the extended spin state lifetime to produce gains in measurement signal-to-noise (SNR) by two orders of magnitude. 

Building on seminal contributions from Elleman \I{et al}~\cite{Rhim76,Rhim78}, our protocol preserves against qubit decay by a two-pronged strategy: $\I{(i)}$ engineering the dipolar spin Hamiltonian $\mH_I\equiv \mH_{dd}$, on average, to commute with the initially prepared transverse qubit state, \I{i.e.} $\lsb \xr_I,\ov{\mH}_{dd}\rsb \rt 0$; and $\I{(ii)}$ by concurrently suppressing the dephasing noise on the qubits, i.e. $\ov{\mH}_z\rt 0$. This is accomplished by requiring only global rotations to be applied to the qubits~\cite{Ajoy13l}. The protocol itself can be viewed as a perturbation on the familiar Carr-Purcell-Meiboom-Gill (CPMG) dynamical decoupling~\cite{Carr54,Meiboom58}, in a manner that additionally renders it immune to inter-qubit interaction. Suitably modified, the qubit preservation strategy here may possibly be harnessed as a spectroscopy tool to selectively unravel dipolar contributions to qubit decay. Indeed, by systematically varying $\mH_{dd}$, we observe that the decay times extend approximately proportional to the relative strength of the interqubit interaction. While demonstrated here for dipolar $\Cs$ qubits in a bulk solid, to the extent that this method relies on coherent averaging of interspin interactions during DD, extensions are applicable to other systems, particularly electronic qubits~\cite{Ryan10,Pla12}, polar molecules~\cite{Ni08,Park17} and Rydberg atoms~\cite{Levine218}. 

\begin{figure}[t]
  \centering
  {\includegraphics[width=0.49\textwidth]{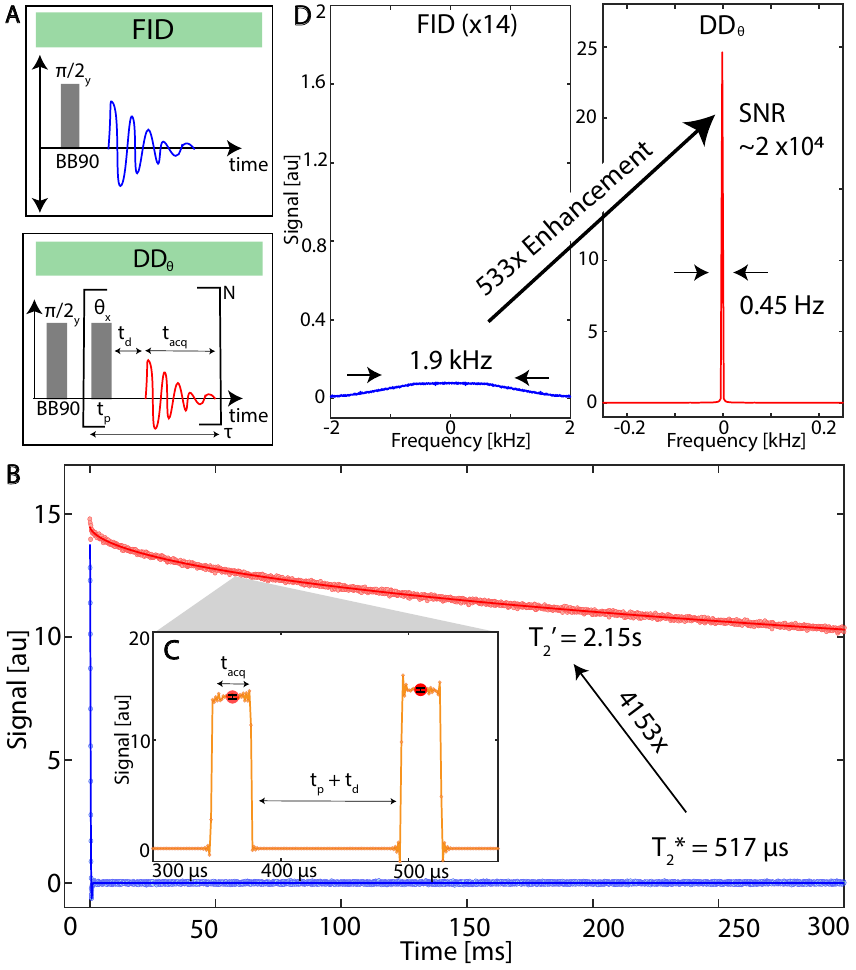}}
  \caption{\T{Transverse state preservation in interacting $\Cs$ spins.} (A) \I{Protocol}. Panels show conventional FID acquisition (\I{top}), and proposed multi-pulse protocol (DD$_{\xt})$ (\I{bottom}), consisting of a train of $N$-equally spaced pulses with flip angle $\xt$. Spins are interrogated in the $\tacq$ periods between the pulses. $t_d$ is a dead time to allow for amplifier switching, and $\qt$ denotes the total cycle period. A BB90 composite pulse provides the $\pi/2$ pulse. (B) \I{Results} in a 10\% enriched crystal. Points denote single-shot data acquired with 20s of optical pumping. Conventional FID decays in $T_2^{\ast}=$517$\mu$s (blue), while the protocol with $\xtopt=$218$^{\circ}$ preserves transverse states for $T_2'=$2.15s (red). Solid lines are fits to stretched exponential decays. (C) \I{Inset:} data collection strategy. 
  The points of the main decay curve (red, with error bars shown) are the medians from the readout signal (orange, sampled at 1MHz) within acquisition windows $\tacq=$32$\mu$s.
 (D) \I{Line narrowing and SNR gains}. Fourier transforms demonstrate line narrowing by $\sim$4200 and an associated $>$500x signal gain. Left panel is zoomed 14-fold. Single-shot SNR in the right panel exceeds $2\zt 10^4$.}
	\zfl{fig2}
\end{figure}

\begin{figure*}[t]
  \centering
  {\includegraphics[width=0.95\textwidth]{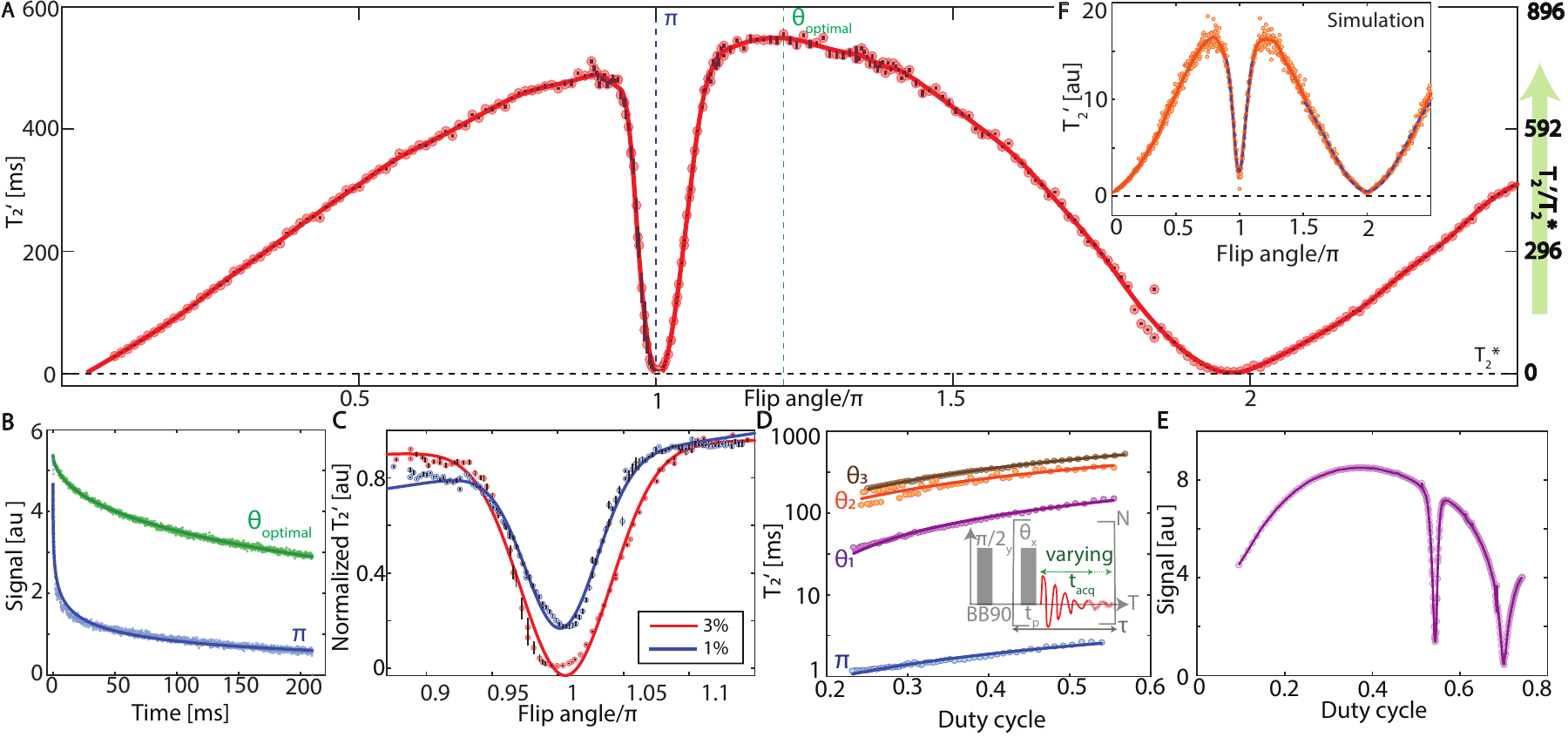}}
  \caption{\T{Origins of state preservation.} (A) \I{Variation of $T_2'$ with flip angle} $\xt$ for a 3\% $\Cs$ enriched sample. Points and error bars are extracted from stretched exponential fits. Solid line is a guide to the eye. Two strong lifetime dips around $\xt=\{\pi,2\pi\}$ are observable.  This data are taken with single-shot (20s) optical pumping, and with total acquisition time $T=N\qt=$200ms and $\tacq$=32$\mu$s. Dashed horizontal line denotes $T_2^{\ast}\app$0.67ms.	Right axis shows the extension ratios $\labs T_2'/T_2^{\ast}\rabs$. Optimal flip-angle $\xtopt$ delivers a lifetime extension by 821-fold (\ztr{table}). (B) Decays at $\xt=\{\pi,\xtopt\}$, demonstrating non-optimality of CPMG control. Solid lines are stretched exponential fits. (C) \I{Widths} of dip-feature at $\xt\app\pi$ for samples with $\eta=\{1\%,3\%\}$, shows an increase with enrichment (see \zfr{fig4}). (D) \I{Variation with duty cycle}. Observed $T_2'$ times on a log scale for varying $\tacq$ (\I{inset}) at fixed values of $\xt$. We display representative values $\xt=\pi$, and $\xt=\{\xt_1,\xt_2,\xt_3\}=\{2.24\pi,1.06\pi,1.16\pi\}$. Solid lines are a linear fit to resulting duty cycle $\eta_d$. (E) \I{Total measured signal} corresponding to (A) demonstrates the interplay between maximizing data acquisition windows ($\eta_d\rt 0$) and lifetime extension ($\eta_d\rt 1$). (F) \I{Simulations} on a $N_s= 6$ spin dipolar network show qualitative agreement. Dashed lines are Gaussian fits to features at $\xt=\{\pi,2\pi\}$. }
\zfl{fig3}
\end{figure*}

\T{\I{System and experimental setup}} -- We consider $\Cs$ nuclear spins (\zfr{fig1}A) in CVD fabricated diamond crystals at high magnetic field, here $B_0=$7T. The $\Cs$ nuclei are prepared in a transverse state denoted by the density matrix, $\xr_I \app \id + \vxe_{\R{pol}} I_x$, where $I_x=\sum_j I_{jx}$, $I$ are spin-$\fr{1}{2}$ Pauli matrices, and $\vxe_{\R{pol}}(\ll 1)$ the polarization level. Evolution under the Hamiltonian
$\mH = \mH_z + \mH_{dd}$ leads to complex many-body dynamics manifesting in the typical $\Cs$ NMR free indication decay (FID~\cite{Lowe59,Baum85}) in $T_2^{\ast}\lesssim$1ms. The $\Cs$ nuclei are dipolar coupled (dashed lines in \zfr{fig1}A) with $\mH_{dd} = \sum_{j<k} d_{jk}^{\CC}(3I_{jz}I_{kz} - \vec{I_j}\cdot\vec{I_k})$, with a coupling strength,
$
d_{jk}^{\CC} = \fr{\mu_0}{4\pi}\hbar\xg_n^2(3\cos^2\xt_{jk}-1)\fr{1}{r_{jk}^3}
$
where, $\xg_n$=10.7MHz/T is the gyromagnetic ratio, and $\xt_{jk}=\cos^{-1}\lb \fr{\T{r}_{jk}\cdot\T{B}_0}{r_{jk}B_0}\rb$ is the angle of the interspin vector $\T{r}_{jk}$ to the magnetic field. If $\eta$ denotes the $\Cs$ enrichment level, the $\Cs$ spin density is $\sim$0.92$\eta$/nm$^3$, and the median dipolar coupling scales $\expec{d_{jk}^{\CC}}\sim\eta^{1/2}$, with a strength $\app$1kHz at natural abundance~\cite{Ajoy19relax}. The spins additionally experience dephasing interactions, $\mH_z = \sum_j c_jI_{jz}$, primarily from the lattice bath of paramagnetic impurities (P1 centers). Here $\expec{c_j}\sim\expec{A_{zz}^2}^{1/2}$ is set by the mean longitudinal P1-$\Cs$ hyperfine interaction~\cite{Reynhardt00}; at typical 20ppm P1 concentrations, $\expec{A_{zz}^2}\app$0.4[kHz]$^2$. 

The diamond samples are doped with optically polarizable electronic defects, \NV centers~\cite{Jelezko06}, at concentration $P_e\sim$1-10ppm. The inter-NV spacing $\expec{r_{\R{NV}}}\app\lb 3/4\pi \ln 2\rb^{1/3}N_e^{-1/3}$, where $N_e=(4\zt 10^{-6}P_e)/a^3$[m$^{-3}$] is the inverse volume concentration, and $a$=0.35nm the diamond lattice spacing~\cite{Reynhardt03a}. Typically, $\expec{r_{\R{NV}}}\app$12nm at 1ppm concentration. The NV centers hyperfine couple to the $\Cs$ nuclei and can transfer their optically induced polarization to them. Such \I{hyperpolarized} $\Cs$ nuclei can be probed inductively with high SNR (see \zfr{fig1}B). We use a recent dynamic nuclear polarization (DNP) approach involving chirped microwave (MW) excitation at low-field ($\Bp\sim$38mT) under continuous optical pumping (520nm, 45mW/mm$^3$)~\cite{Ajoy17}. Ratchet-like polarization transfer is set up by exciting a pair of NV-$\Cs$ Landau-Zener transitions in the rotating frame, forcing either adiabatic or diabatic traversals conditional on the nuclear spin state~\cite{Zangara18}. Spin diffusion approximately homogenizes $\Cs$ polarization in the $\fr{\expec{r_{\R{NV}}}}{2}$ radii spherical regions between neighboring NVs. Optical pumping by 40s can result in $\Cs$ polarization levels $\vxe_{\R{pol}}\app$1.2\% (see \zfr{fig1}B). This corresponds to a boost in polarization over the 7T Boltzmann value by $\vxe\app$998, and accelerates $\Cs$ inductive detection by $\vxe^2\lsb T_1(B_0)/T_1(\Bp)\rsb^2\gtrsim 10^9$ over conventional 7T NMR. This results in high single-shot $\Cs$ NMR SNRs $\app 10^3$ for typical $\sim$17mg diamond crystals.

\T{\I{Transverse state preservation}} -- The proposed control sequence (see \zfr{fig2}A) consists of a train of $N$ equally spaced pulses with flip-angle $\xt$; in general, $\xt\neq\pi$.  Indeed, for optimal performance, $\xtopt =\pi+\xD\xt$, where $\xD\xt/\pi\lesssim 1/4$. The $\Cs$ nuclei are interrogated in each period between pulses (less a dead time $t_d$) for duration $\tacq$; given pulse widths $t_p$, we typically employ duty cycles $\eta_d=t_p/\qt\sim$0.5, where $\qt$ is the total cycle period (\zfr{fig2}A). We refer to the sequence (\zfr{fig2}A) as DD$_{\xt}$ due to similarities with dynamical decoupling; several familiar sequences including CPMG decoupling~\cite{Carr54,Meiboom58} ($\xt=\pi$), Waugh-Ostroff trains ($\xt=\pi/2$)~\cite{Ostroff66}, and \I{cw} spin-locking~\cite{Hartman62} ($\tacq=0$) are identifiable as special cases. The pulsed spin-locking approach of Elleman \I{et al}~\cite{Rhim76,Rhim78} dealt with this protocol largely in the limit of small $\xt$ and purely dipolar interactions. As we demonstrate, extensions of this methodology to large $\xt(=\pi+\xD\xt)$  renders it suitable for dynamical decoupling when \I{both} dephasing and dipolar interactions are operational on the qubits. Rapid hyperpolarization and the high SNR in our experiments allow study of sequence performance (e.g. dependence on $\xt$) with about two orders of magnitude higher density of points than prior NMR experiments. This more clearly unravels optimum regimes for the applied quantum control.

\zfr{fig2}B shows representative results for a $\eta=$10\% sample. In contrast to normal FID decay $T_2^{\ast}=517\pm3\mu$s, the multi-pulse protocol with flip-angle $\xt=$218$^{\circ}$ leads to long-time state protection to $T_2'=2.147\pm$0.022s, corresponding to an extension factor $\left|T_2'/T_2^{\ast}\right|\app$4159. Here, the duty cycle $\eta_d=$0.6, with acquisition windows $\tacq=$32$\mu$s. Since $\tacq\ll \{T_2^{\ast},T_2'\}$, points from each $\tacq$ window are averaged over (\I{decimated}, see \zfr{fig2}C) to generate the decay curves sampled at $\qt^{-1}$. We observe qualitatively similar extensions in $\Cs$ state lifetimes for diamond crystals at multiple enrichment values (\zfr{fig4}), as well as powders.

The extended spin interrogation leads to $\Cs$ NMR line narrowing and consequent SNR boosts. \zfr{fig2}D shows Fourier transforms corresponding to data in \zfr{fig2}B, wherefrom we estimate an SNR gain by $\mS\app$533. The SNR enhancement over conventional FID detection, in general, lies between $\lsb(1-\eta_d)T_2'/T_2^{\ast}\rsb^{1/2}<\mS<(1-\eta_d)T_2'/T_2^{\ast}$. The linear scaling upper bound arises when both FID and DD$_{\xt}$ curves are acquired for long periods $T\geq T_2'$ (a situation pertinent to quantum sensing applications). The lower bound ($\app$46 in \zfr{fig2}B) saturates when the curves are acquired to their respective decay constants $\{T_2^{\ast},T_2'\}$. The overall SNR consequent to DD$_{\xt}$ acquisition is rather large (see \zfr{fig2}D) $\app 2\zt 10^4$ in 40s of optical pumping for a 17mg diamond, and is limited by finite memory capacity (250k complex points)~\cite{SOM}. 
Furthermore, the low filling-factor ($f\app$1/500) in our homemade NMR probe~\cite{Ajoyinstrument18} degrades detection SNR, and reduces acquisition duty cycle $(1-\eta_d)$ for a given $\xt$ due to limited $\Cs$ Rabi frequency ($\xO\app$11.4kHz). Nonetheless we are able to attain an effective acceleration by $\lsb\vxe\fr{T_1(B_0)}{T_1(\Bp)}\rsb^2\fr{(1-\eta_d)T_2'}{T_2^{\ast}}\sim 10^{11}$ over conventional 7T NMR detection through a combination of hyperpolarization and extended interrogation.

\begin{figure*}[t]
  \centering
  {\includegraphics[width=0.95\textwidth]{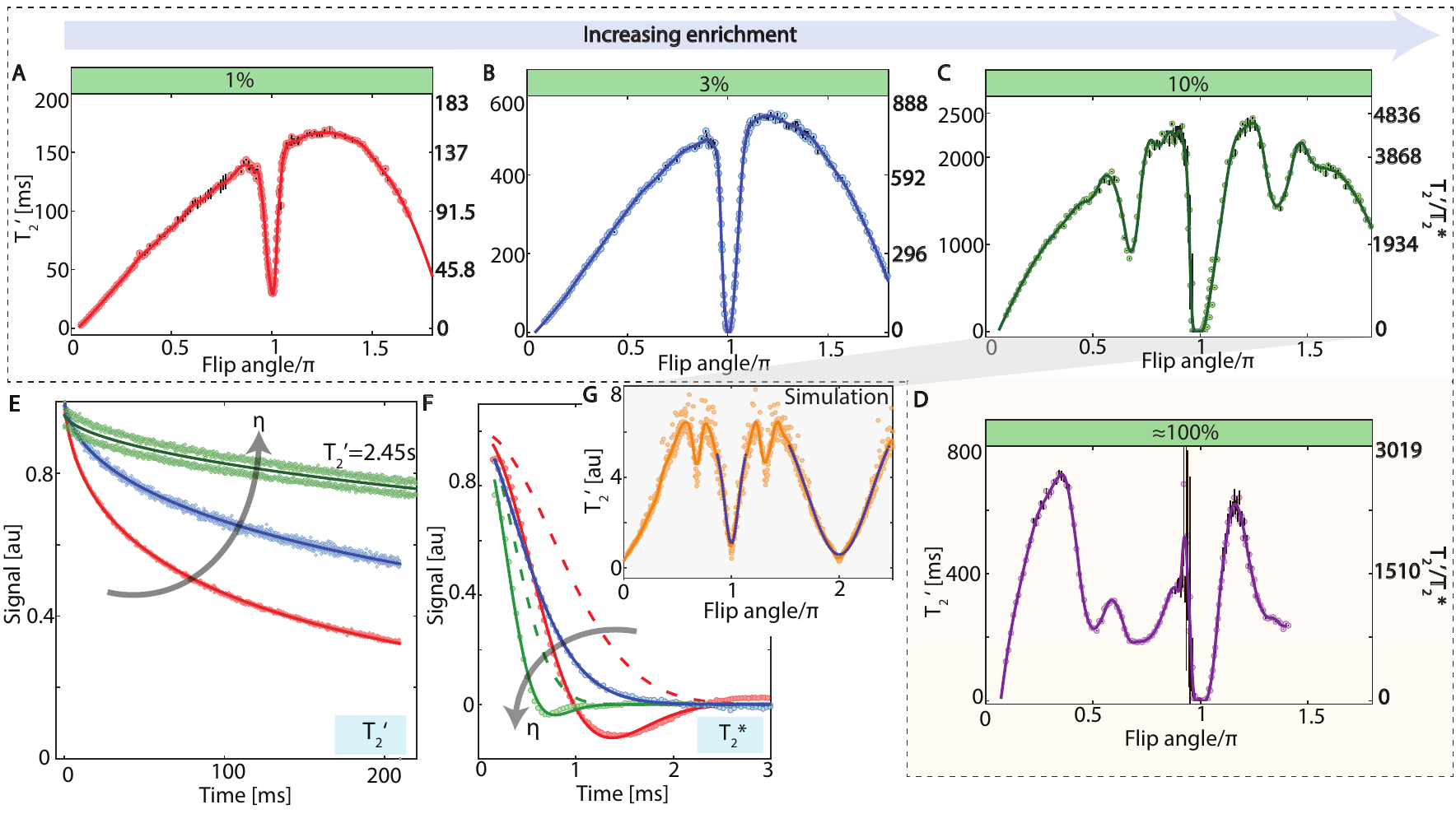}}
  \caption{\T{Variation with $\Cs$ enrichment}. (A-D) Measured $T_2'(\xt)$ values for crystals with $\eta = \{$1\% (red), 3\% (blue), 10\% (green), 100\% (purple)$\}$ enrichment. Points and error bars are extracted from $1/e$ intersections of stretched exponential fits. Solid lines are spline interpolated guides. Right axes (bold) denote $\labs T_2'/T_2^{\ast}\rabs$ extension factors, indicating improvement with enrichment. (E)  \I{Single-shot $T_2'$ decays} show lifetime improvement with $\eta$ up to 10\%  (arrow, see \ztr{table}). Displayed are decays at $\xtopt$ with respective stretched exponential fits. (F) \I{FID decays} show expected decreases in $T_2^{\ast}$ with increasing enrichment $\eta$ (arrow). Dashed lines indicate exponential decay envelopes.  (G) \I{Simulations} with $\|\mH_{dd}\|=4\|\mH_{z}\|$, qualitatively recover the experimental features in \zfr{fig4}C.}
\zfl{fig4}
\end{figure*}

To unravel physical origins of qubit state preservation, in \zfr{fig3}A we measure $T_2'(\xt)$ times for varying flip angles $\xt$ in a 3\% enriched sample. The dashed line denotes the sample $T_2^{\ast}\app$0.67ms, and the resulting $\left|\fr{T_2'}{T_2^{\ast}}\right|$ extension factors ($\gtrsim$850) are elucidated on the right axes. Here the total acquisition time, dead time, and acquisition periods are fixed to $T=$200ms, $t_d=6\mu$s, and $\tacq=$32$\mu$s respectively. This translates to an increasing duty cycle $\eta_d=\lb 1 + 2\pi\xO\tacq/\xt\rb^{-1}$, with increasing $\xt$. For small $\xt$, $\eta_d\app \xt/(2\pi\xO\tacq)$ scales linearly with flip-angle. Experimentally, we observe an approximately linear increase in $T_2'$ in this regime, tracking the increasing duty cycle. Prominent in the data (\zfr{fig3}A) is a sharp drop in the $T_2'$ lifetimes at $\xt=\pi$ (CPMG condition), and a much broader secondary drop around $\xt=2\pi$. Optimal state preservation manifests at $\xtopt=\pi+\xD\xt$, with $\xD\xt/\pi\app$0.22, indicating that even a few degrees difference in the flip angles can provide a significant difference (here 280-fold) in the decay times. These results indicate that while long CPMG spin echoes have been observed for solids previously~\cite{Li07,Li08,Ridge14}, this is potentially suboptimal for spin networks with significant interspin coupling, mainly due to inefficacy of counteracting $\mH_{dd}$ as discussed below. 

\zfr{fig3}B shows characteristic decays at $\xtopt$ and $\xt = \pi$ (CPMG) and indicates a markedly stretched exponential behavior in the latter. We note that while CPMG decay is far from optimal ($T_2'$=6.3ms), it is still an order of magnitude slower than the conventional FID. \zfr{fig3}C compares the normalized $T_2'$ dips in the region near $\xt=\pi$ for 1\% and 3\% $\Cs$ enriched samples. The Gaussian widths are broader in the latter case, indicating a relationship between this feature and $\|\mH_{dd}\|$. While \zfr{fig3}A considers the case of fixed $\tacq$, we evaluate in \zfr{fig3}D performance with varying $\tacq$ (hence $\eta_d$) for fixed values of $\xt$ (see inset). Data here is shown a log scale for clarity, and solid lines are linear fits. The excellent linear agreement indicates that $T_2'\propto\xn_d$ track the pulse duty cycle for all $\xt$. Indeed, at $\eta_d\rt 1$ and small $\xt$, the lifetime $T_2'(\xt)\rt T_{1\rho}(\xO)$, where $T_{1\rho}$ is the full spin-locking lifetime. Extrapolating from \zfr{fig3}D, we estimate $T_{1\rho}\app$1s. This brings forth the trade-off inherent in the sequence between protecting the spin state for longer periods (large $\eta_d$) and opening $\tacq$ windows where the spins can be interrogated (small $\eta_d$). The effect of these opposite regimes is illustrated in \zfr{fig3}E, where we display the measured signal strength corresponding to data in \zfr{fig3}A. Indeed, assuming negligible measurement dead time, the total acquired signal scales $\approxprop\xn_d(1-\xn_d)$, and the maximum signal is obtained at $\xn_d\app$0.5. We note finally that our sequence construction in \zfr{fig2}A is likely optimum with respect to NMR readout SNR. Signal acquisition can be carried out after every applied pulse, since the minimal repeated block is just a single pulse, leading to an optimally high readout duty cycle compared to more complex multipulse approaches.

\T{\I{Theory}} -- To elucidate sequence operation, we assume pulses are applied on resonance at $\xo_0=\xg_nB_0$ and are $\xd$-like. In the rotating frame at $\xo_0$, $\mH=(\mH_{dd} + \mH_z)$ is the system Hamiltonian, and sequence action corresponds to the propagator, $U(N\qt) = [\exp(i\xt I_x)\exp(i\qt\mH)]^N$, where $\qt=(t_p+\tacq + t_d)$. We assume negligible evolution under the pulses, reasonable when $\|\mH\|t_p\ll1$, and applicable here since $\|\mH\|\lesssim$5kHz and $t_p\lesssim$60$\mu$s. Starting with an initial state $\xr_I\sim I_x$, the qubit survival probability is $F(N\qt)=\fr{1}{2}\Tr{\xr_I^{\dg}U(N\qt)^{\dg}\xr_IU(N\qt)}\approxprop e^{-N\qt/T_2'}$. Assuming for simplicity $(N+1)$ pulses, the propagator can be expressed as, $U =  e^{i(N+1)\xt I_x}\prod_{n=1}^{N} \exp(i\qt\mH^{(n)})$, where the toggling frame Hamiltonians, $\mH^{(n)}= e^{in\xt I_x}\mH e^{-in\xt I_x} = \mH_{dd}^{(n)} + \mH_z^{(n)}$~\cite{SOM}. The evolution can be considered driven by an effective Hamiltonian $U\app \exp(iN\qt\ov{\mH})$, given by a Magnus expansion~\cite{Magnus54}, $\ov{\mH} = \ov{\mH}^{(0)} + \ov{\mH}^{(1)} +\cdots$. The leading (zeroth) order term $\ov{\mH}^{(0)}=\fr{1}{N}\sum_n \mH^{(n)}$ can approximately capture system dynamics when $\qt\|\mH\|\ll$1 and the 
expansion converges; this is the predominant regime in our experiments. Considering first the dipolar Hamiltonian we have ~\cite{SOM},
\bea
\ov{\mH}^{(0)}_{dd} &=& \sum_{j<k}d_{jk}^{\CC}\lb \fr{3}{2}\lsb \mH_{\R{ff}} + \mG(\xt)\lb\mH_{\R{dq}}\cos(N+1)\xt \right.\right.\right.\non\\
&+& \left.\left.\left.\wt{\mH}_{\R{ff}}\sin(N+1)\xt\rb\rsb -\vec{I_j}\cdot\vec{I_k}\rb\:,
\zl{Hdd}
\eea
with the flip-flop and double-quantum Hamiltonians,
$\mH_{\R{ff,dq}} = I_{jz}I_{kz}\pm I_{jy}I_{ky}$, tilted flip-flop $\wt{\mH}_{\R{ff}}=I_{jz}I_{ky}+ I_{jy}I_{kz}$, and 
the \I{grating} function, $\mG(\xt) = \fr{1}{N}\lb\fr{\sin N\xt}{\sin\xt}\rb$~\cite{Ajoy13l}. This function resembles an optical diffraction grating, with $\mG(\xt)\rt 0$ for $\xt\neq n\pi$, with peaks $\mG(n\pi) = 1$, and with a linewidth falling rapidly $\sim 1/N$~\cite{Ajoy13l}. It acts as a {filter}, engineering the effective Hamiltonian in \zr{Hdd}; in the large $N$ limit, for $\xt\neq n\pi$ we have,
\beq
\ov{\mH}^{(0)}_{dd}\app \sum_{j<k}d_{jk}^{\CC}\lb \fr{3}{2}\mH_{\R{ff}}-\vec{I_j}\cdot\vec{I_k}\rb.
\eeq 
State preservation follows since the average Hamiltonian commutes with the initial state, $[\ov{\mH}^{(0)}_{dd},\xr_I]=0$, locking the spins against decay in the rotating frame. Hamiltonian filtering is fastest for $\xt=\pi/2$ (Waugh-Ostroff condition), since $\mG(\pi/2)$=0. Even otherwise at large $N$ ($\gtrsim$2000 in our experiments), the filter converges effectively for arbitrary $\xt(\neq n\pi)$. Defining a \I{convergence} length $L$ as the number of pulses such that the phase accumulated by terms in $\mH_{dd}^{(n)}$ approaches $2\pi$, we have $L\sim |\sin^{-1}[\sin \xt]|$. At $\xt=\pi$ (CPMG condition), on the other hand, the full dipolar coupling remains operational on the spins, $\ov{\mH}_{dd}=\mH_{dd}$ (see~\cite{SOM}), yielding the rapid decay at $\xt/\pi \app 1$ in \zfr{fig3}A. The \I{approach} to the recoupling at $\xt=\pi$ scales with $\|\mH_{dd}\|$ (see \zfr{fig3}C), and is evidenced also in numerical simulations ~\cite{SOM}.

We note importantly that the condition $[\ov{\mH}^{(0)}_{dd},\xr_I]=0$ is simpler to accomplish than traditional \I{decoupling} approaches~\cite{Waugh68, Frey12} that largely try to engineer $\ov{\mH}^{(0)}_{dd}\rt 0$. The latter however comes at a higher cost of pulse sequence complexity, involving multiple phase switches and/or flip angles. In contrast, the sequence in \zfr{fig2}A just involves the repeated application of a single $\xt$-pulse, is simple to implement experimentally, and manifests as minor perturbation of conventional CPMG DD.

A similar approach can be taken to consider the action of the sequence on dephasing interactions, leading to the average Hamiltonian,
\beq
\ov{\mH}^{(0)}_{z}=\sum_jc_{j}\mG(\xt/2)\lsb I_z\cos(N+1)\xt/2 +I_y\sin(N+1)\xt/2\rsb\:.
\eeq
Since $\labs\mG(\xt/2)\rabs\rt 0$ for $\xt\neq 2n\pi$, the dephasing Hamiltonian is then \I{decoupled}, \I{i.e.} $\ov{\mH}^{(0)}_{z}\app 0$. Hamiltonian filtering is most rapid for $\xt=\pi$ (CPMG condition), with a convergence length $L\sim |\sin^{-1}[\sin \xt/2]|$. On the other hand $\xt=2\pi$ leads to the \I{full} Hamiltonian $\ov{\mH}=(\mH_{dd} + \mH_z)$ operational on the spins, yielding the prominent drop at $\xt/\pi \app 2$ in \zfr{fig3}A.

In summary, state protection arises by engineering Hamiltonian $\mH$ such that the initial state is rendered immune to decay from both dephasing as well as inter-qubit interactions. The optimal flip angle $\xtopt$ arises from an interplay between optimal conditions (CPMG and Waugh-Ostroff) for both these limiting cases~\cite{SOM}. We note that while we considered only zeroth-order average average Hamiltonian above, the first order terms can be calculated similarly and shown to have similar Hamiltonian filtering properties (see \cite{SOM}). For instance, $\ov{\mH}^{(1)}_{dd} = -\frac{i\qt}{2N}\sum_{n=1}^{N-1} \sum_{\ell=n+1}^{N}  [\mH^{(n)}_{dd},\mH^{(\ell)}_{dd}]$ filters out except for $\xt=n\pi$, leading to $\ov{\mH}^{(1)}_{dd}\rt 0$ for most values of $\xt$.

Numerical simulations affirm this model of sequence operation (\zfr{fig3}F). We consider small networks of $N_s=6$ $\Cs$ spins arranged in a diamond lattice. Dipolar couplings are calculated from relative spin positions, and a random dephasing field a
\begin{flushright}

\end{flushright}
pplied at every spin site (see~\cite{SOM}). Under $\xd$-like $\xt$ flip-angle pulses, decay times are estimated from the survival probabilities as $T_2'(\xt)\app -N\qt/\log(F(N\qt))$~\cite{SOM}. Averaging over 50 network manifestations we obtain $T_2'(\xt)$ profiles (\zfr{fig3}F) that qualitatively match the experimental data. Tuning the relative matrix norms, $\|\mH_{dd}\|/\|\mH_{z}\|$, allows the study of the individual contributions to the decay. In the weak coupling limit $\|\mH_{dd}\|\lesssim\|\mH_{z}\|$, dipolar couplings dominate the width of the features near $\xt=\pi$, while dephasing dominates those near $2\pi$~\cite{SOM}. $\xtopt$ arises as an interplay between these two decay widths. Increasing $N$ improves the state preservation lifetimes but does not narrow the width of the $T_2'(\xt)$ profiles, which instead depends only on the Hamiltonian strengths.

\begin{table}[t]
\begin{tabular}{|c|c|c|c|c|c|>{\columncolor{Gray}} c|}
\hline
\mc{1}{\T{$\Cs$ Enrichment}} & \T{$T_1(B_0)$ [s]}~\cite{Ajoy19relax}& \T{$T_2^{\ast} [\mu$s]}   & \T{$T_2'$ [ms]} & \T{$\xtopt$ [deg]} & \T{$\left|\fr{T_2'}{T_2^{\ast}}\right|$} \\ \hline
\hline
1 \%        & 1167$\pm$ 14          & 1092$\pm$ 4  & 169$\pm$ 3.9           & 232 $\pm$ 18       & 154                                                         \\ \hline
3\%        & 1028 $\pm$ 171           & 676$\pm$ 195   & 555$\pm$7.5           & 218$\pm$ 16       & 821                                                         \\ \hline
10\%         & 122$\pm$3             & 517$\pm$ 3   & 2447$\pm$14           &  224$\pm$ 4     & 4733                                                         \\ \hline
$\app$ 100\%         & 42$\pm$2         & 265$\pm$ 30  & 639$\pm$ 12            & 209 $\pm$ 10 &2411                                                      \\ \hline
\end{tabular}
\caption{\T{Summary} of measured $T_2'$ lifetimes of samples with varying $\Cs$ enrichment (see \zfr{fig3} and \zfr{fig4}). See ~\cite{SOM} for error bar calculations.}
\ztl{table}
\end{table}

\T{\I{Ultralong preservation with enrichment}} -- We observe a surprising increase in $T_2'$ values with increasing $\Cs$ enrichment $\eta$ (\zfr{fig4}A-D). While the $T_2'(\xt)$ profiles are qualitatively similar across samples, the increase in state lifetime is evident in the relative scaling of the $y$-axes in \zfr{fig4}A-D, and from $|T_2'/T_2^{\ast}|$ elucidated in the right axes. For perspective, \zfr{fig4}F illustrates the FIDs, wherein we define $T_2^{\ast}$ values from exponential decay envelopes (dashed lines). The decrease with enrichment (gray arrow) is expected from the scaling of the mean dipolar coupling $d_{jk}^{\CC}$.
Other sample parameters, notably $T_1(B_0)$ (\ztr{table}), degrade similarly with enrichment~\cite{Ajoy19relax}. In comparison, $T_2'(\xtopt)$ (see \zfr{fig4}E) proceeds counter to this trend. While a physical basis is beyond the scope of this manuscript, we speculate a ``localization'' effect impedes state decay at higher $\Cs$ enrichment. The increasingly complex $T_2'(\xt)$ features in the highly enriched samples are challenging to predict analytically except at $\xt=\pi$. Numerical simulations (\zfr{fig4}G) in the limit of strong coupling $\|\mH_{dd}\|\app 4\|\mH_{z}\|$ reproduce features at $\xt=\{2\pi/3,4\pi/3\}$ in \zfr{fig4}C. We will consider a more detailed exposition in future work. 

\T{\I{Outlook}} -- With a view toward improving {both} qubit state preservation, as well as total inductive SNR, the current protocol can be improved through several means. Restrictions stemming from the poor sample filling-factors ($f\app$1/500) in our current experiments can be circumvented by the use of surface coils patterned on the diamond ($f\rt$0.5). The resulting {transreceiver} gains in \I{(i)} pulse Rabi frequency, allowing greater acquisition duty cycles for a given $\xt$, and \I{(ii)} inductively measured SNR, both independently scale $\approxprop f^{1/2}$~\cite{Hoult1978}.
Furthermore, data here was acquired at $\lesssim$1MS/s leading to $<$40 points per $\tacq$ window. Faster digitization or boxcar averaging will boost SNR by at least an order of magnitude.  With these improvements, single-shot SNR could approach $\gtrsim 10^7$ for a 1mg diamond sample. These \I{twin gains} in high-SNR and narrow linewidth detection of hyperpolarized $\Cs$ nuclei may open opportunities for their use in hyperpolarized imaging~\cite{Lv19} and in quantum sensing as gyroscopes~\cite{Ajoy12g,Ledbetter12} and magnetometers~\cite{Degen17}. The latter is based on discerning, via changes in the coherence time $T_2'$, weak fluctuating magnetic fields that are matched in frequency to a resonant period of the pulse delay $\qt$.

\T{\I{Conclusions}} -- We have demonstrated a signal-boosted dynamical decoupling amenable to the regime where qubits are immersed in a dephasing environment, and wherein inter-qubit interaction is comparably non-negligible. For dipolar coupled $\Cs$ nuclei in diamond we demonstrated extensions of qubit state preservation lifetimes and SNR measurement gains, both by about three orders of magnitude. We anticipate applications of a similar approach in other interacting qubit systems, including in polar molecule and Rydberg atom platforms.

\T{\I{Acknowledgments}} -- We gratefully acknowledge discussions with S. Bhave, D. Bugarth and D. Wemmer. A.A acknowledges support from ONR under N00014-20-1-2806. Partial support was provide by DOE under DE-AC02-05CH11231 and NSF GOALI under 1903803. A.M.S. acknowledges support from CNPq, FAPERJ (grant 203.166/2017) and INCT-IQ. C.A.M acknowledges support from RSC through a FRED Award and the NSF CREST-IDEALS under NSF-HRD-1547830. 

\I{Note added} -- While this work was under review, two additional papers have been recently published on a related subject~\cite{Zhou20, Choi20}.


\bibliography{Biblio}
\bibliographystyle{apsrev4-2}

\pagebreak

\clearpage
\onecolumngrid
\begin{center}
\textbf{\large{\textit{Supplementary Information} \\\smallskip
\bluetitle{Dynamical decoupling in interacting systems: applications to signal-enhanced hyperpolarized readout}}}\\
\hfill \break
\smallskip
A. Ajoy,$^{1,\BRd{\ast}}$ R. Nirodi,$^{1}$ A. Sarkar,$^{1}$ P. Reshetikhin,$^{1}$ E. Druga,$^{1}$ A. Akkiraju,$^{1}$ M. McAllister,$^{1}$ G. Maineri,$^{1}$ \\
S. Le,$^{1}$ A. Lin,$^{1}$ A. M. Souza,$^{2}$ C. A. Meriles,$^{3}$ B. Gilbert,$^{4}$ D. Suter,$^{5}$ J. A. Reimer,$^{6}$ A. Pines$^{1}$\\
\smallskip
\emph{${}^{1}$ {\small Department of Chemistry, University of California, Berkeley, Berkeley, CA 94720, USA.}}\\
\emph{${}^{2}$ {\small Centro Brasileiro de Pesquisas Físicas, Rua Dr. Xavier Sigaud 150, 22290-180 Rio de Janeiro, Rio de Janeiro, Brazil.}}\\
\emph{${}^{3}$ {\small Department of Physics and CUNY-Graduate Center, CUNY-City College of New York, New York, NY 10031, USA.}}\\
\emph{${}^{4}$ {\small Energy Geoscience Division, Lawrence Berkeley National Laboratory, Berkeley, CA 94720, USA.}}\\
\emph{${}^{5}$ {\small Fakult\"{a}t Physik, Technische Universit\"{a}t Dortmund, D-44221 Dortmund, Germany.}}\\
\emph{${}^{6}$ {\small Department of Chemical and Biomolecular Engineering, and Materials Science Division Lawrence Berkeley National Laboratory, University of California, Berkeley, Berkeley, CA 94720, USA.}}
\end{center}

\twocolumngrid

\beginsupplement

\begin{figure}[t]
  \centering
  {\includegraphics[width=0.35\textwidth]{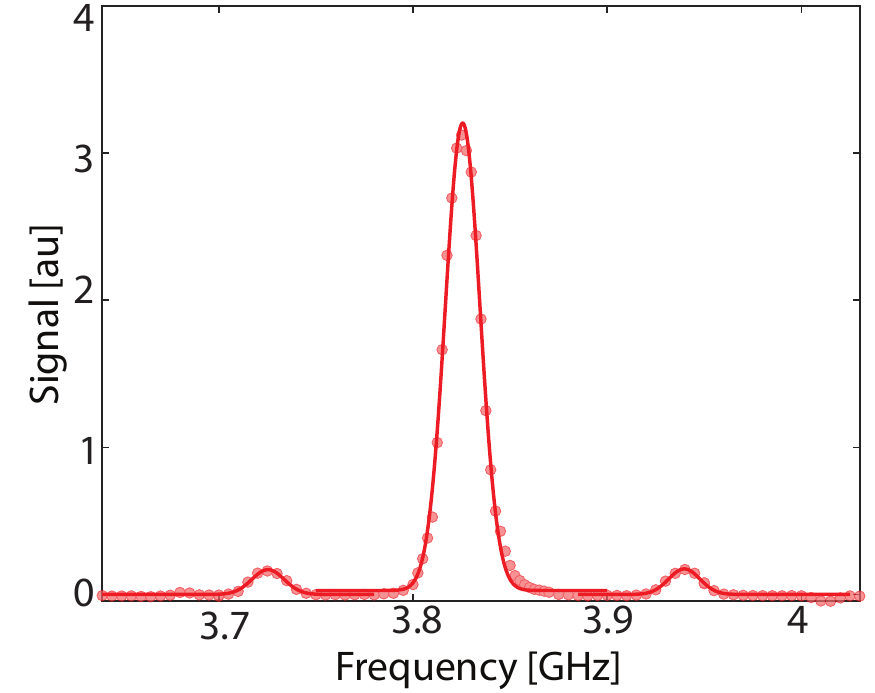}}
  \caption{\T{Experimentally mapped NV ESR spectrum} for the $\eta$=3\% enriched sample. We employ MW sweeps over a 10MHz bandwidth, whose center (abscissas of points) is varied. The resulting hyperpolarized $\Cs$ NMR signal faithfully reflects the underlying NV ESR spectrum. Experiments are performed at $\app$38mT and the samples are placed with [100] axis parallel to the polarization field.  Solid lines are Gaussian fits. In the data in Fig. 3 and Fig. 4 in the main paper, we sweep over the FWHM of the main ESR peak feature.}
\zfl{ESR}
\end{figure}

\begin{figure}[t]
  \centering
  {\includegraphics[width=0.3\textwidth]{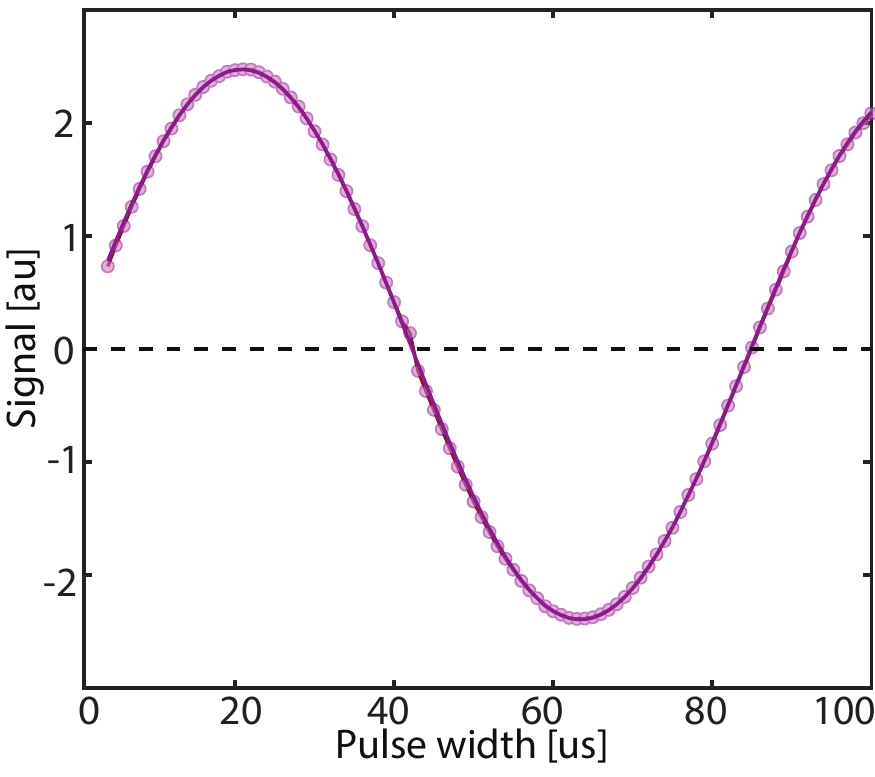}}
  \caption{\T{Exemplary single-shot Rabi oscillation} of $\Cs$ spins in the 10\% enriched sample obtained employing the $DD_{\xt}$ sequence with $\xt=\pi/2$. We measure a Rabi frequency $\xO\app$11.4kHz.}
\zfl{rabi}
\end{figure}

\section{Experimental Apparatus}
Hyperpolarization is accomplished on the crystals employed in this paper by the use of 14 lasers (Lasertack PD-01289). The lasers are fiber coupled (Thorlabs M35L01), resulting in a power output $\app$850mW each, and are arranged in a dome-shaped configuration to ensure uniform sample illumination.  The samples are immersed in water and mounted securely in a 8mm NMR tube. Water here serves as a high thermal capacity heat exchange medium, and is passively cooled by means of a 3D printed Dyson fan delivering high flow-rate air at 1.5$^{\circ}$C. This ensures high-power laser illumination is not associated with significant sample heating.

Laser illumination is accompanied by swept MW irradiation directly synthesized by an arbitrary waveform generator (Tabor SE5082), and upconverted by a mixer (Fairview FMFX1050) with a local oscillator signal produced by a signal generator (Stanford Research Systems SG386). We estimate a total laser power density of $\sim$45mW/mm$^3$, which is sufficient for polarization saturation in the $<$20mg diamond samples considered in this paper. The MWs are amplified by means of a 50W amplifier (Minicircuits ZHL-50W-63+) and delivered by means of a split coil stub antenna, which is tuneless. We estimate a MW transmission of $<$10\%, and a total MW power density $<$10mW/mm$^3$. 

For optimal hyperpolarization we typically choose a MW sweep bandwidth to match that of the NV ESR spectrum ($\sim$30MHz for 1\% enrichment) (see \zfr{ESR}). The single crystal samples are oriented at $\app$54$^{\circ}$ to the main polarization bias field $B_0\sim$38mT, allowing the four NV axes to overlap in frequency. The MW sweeps are set to 600Hz, which we determine to be optimal for the laser power employed~\cite{Ajoy20}. The sweep window bandwidths in general scale with increasing enrichment of the crystals.   The samples are rapidly shuttled ($<$1s) to 7T where the $\Cs$ nuclei in the samples are finally interrogated. NMR spectra are acquired by means of a Varian spectrometer. To evaluate the enhancement factor produced through DNP in our experiments, we evaluate the ratio of the signals obtained through DNP and by conventional 7T NMR (see \zfr{fig1}). We measure typical enhancement values $\vxe>$500 for all samples considered in this work. For more details of hyperpolarization mechanism and the sample shuttling apparatus, we refer the reader to instrumental details in Ref.~\cite{Ajoyinstrument18,Ajoy20}.

We employ a home-built NMR probe for $\Cs$ NMR detection. The antenna is a saddle coil with diameter 11mm and height 13mm cut out of an adhesive copper sheet using a vinyl cutter. We measure a coil quality factor $Q\app$30 at 75MHz.  The need for mechanical clearance for the shuttling process results in non-ideal filling factors ($f\sim 1/500$) for these experiments. We measure a $\Cs$ Rabi frequency of $\xO\app$11.4kHz (see \zfr{rabi}). For more details of the probe and coil construction, we refer the reader to Ref. \cite{Ajoyinstrument18}. Future improvements in filling factor, for instance, through the use of planar coils, can increase the measured SNR by a further one order of magnitude. 

\begin{figure*}[t]
  \centering
  {\includegraphics[width=0.9\textwidth]{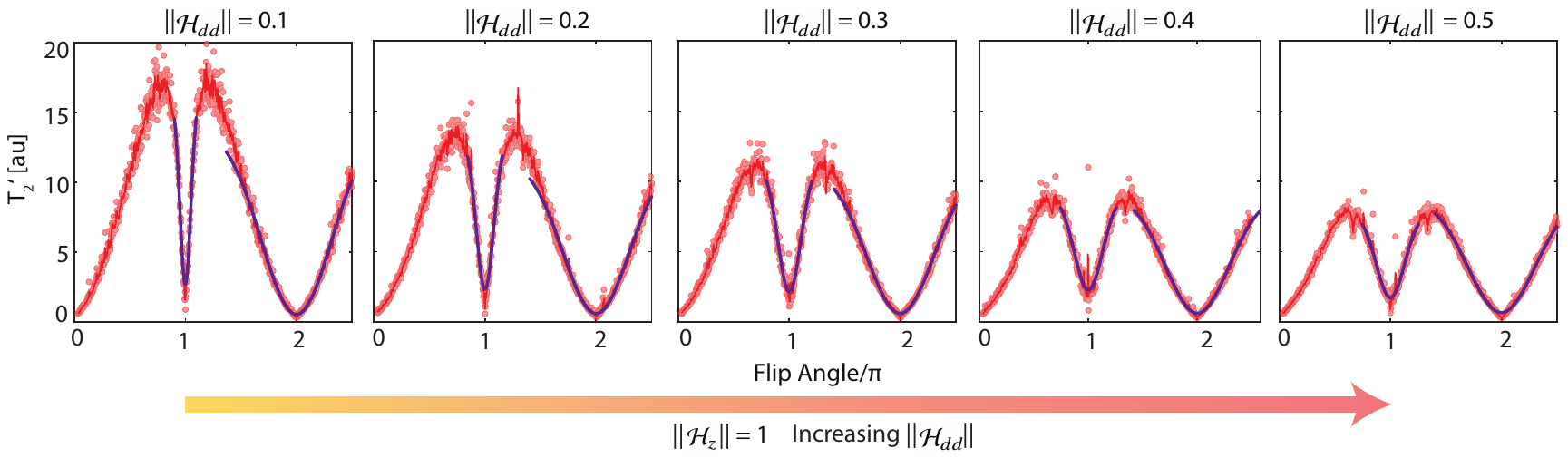}}
  \caption{\T{Simulations indicating the effect of inter-qubit dipolar strength} $\|\mH_{dd}\|$ on $T_2'(\xt)$ profiles. We consider a network of $\Cs$ spins employing $N_s=$6 spins, and average over 50 random manifestations. Here the norm of the inter-spin dipolar coupling $\|\mH_{dd}\|$ is artificially weighted with respect to that of the dephasing interaction $\|\mH_{z}\|$, operating in the weak coupling limit, $\|\mH_{dd}\|<\|\mH_{z}\|$. The blue lines indicate Gaussian fits of the features near $\xt=\{\pi,2\pi\}$. We observe that the dip near $\xt=\pi$ becomes wider with increasing $\mH_{dd}$, while the feature near $\xt=2\pi$ is only weakly affected in comparison.}
\zfl{supp_Hdd}
\end{figure*}

\begin{figure*}[t]
  \centering
  {\includegraphics[width=0.9\textwidth]{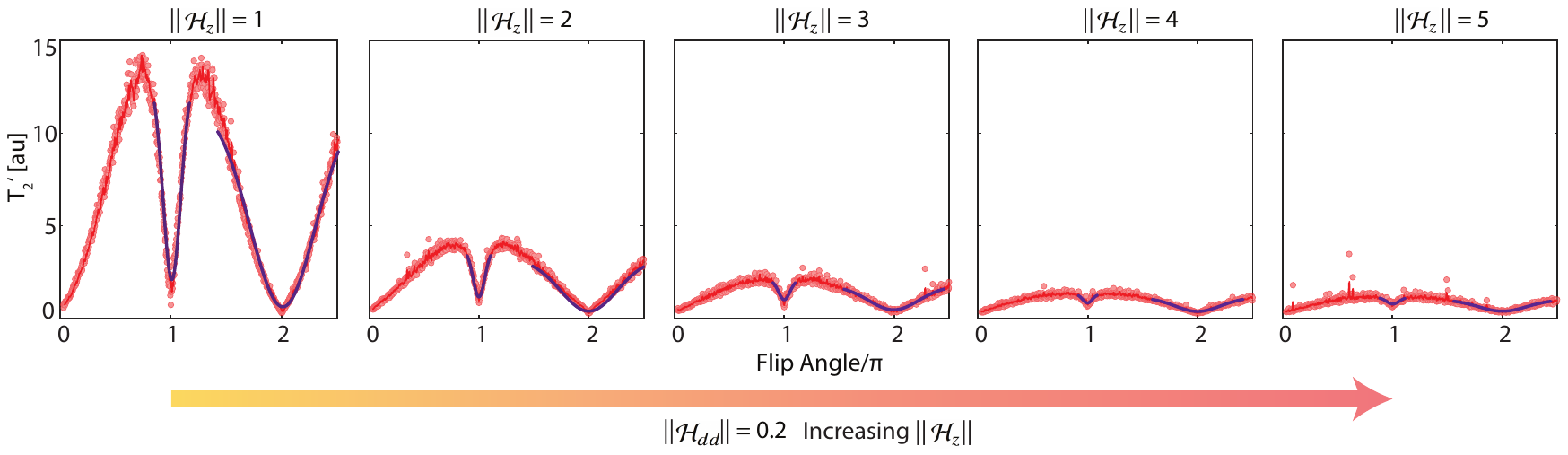}}
  \caption{\T{Simulation showing the effect of dephasing strength} $\|\mH_{z}\|$ on $T_2'(\xt)$ features. Similar to \zfr{supp_Hdd}, we consider a network of $\Cs$ spins employing $N_s=$6 spins. Here the norm of the dipolar interaction $\|\mH_{z}\|$ is artificially weighted with respect to that of the dephasing interaction $\|\mH_{dd}\|$, and we assume that we are in the weak coupling limit. The width of the feature near $\xt=2\pi$ increases with the norm of $\mH_{z}$, while the feature at $\xt=\pi$ is affected negligibly in comparison, suggesting its origin depends solely on the inter-spin dipolar coupling.}
\zfl{supp_Hz}
\end{figure*}

\section{Simulations of sequence operation}

\begin{figure}[t]
  \centering
  {\includegraphics[width=0.45\textwidth]{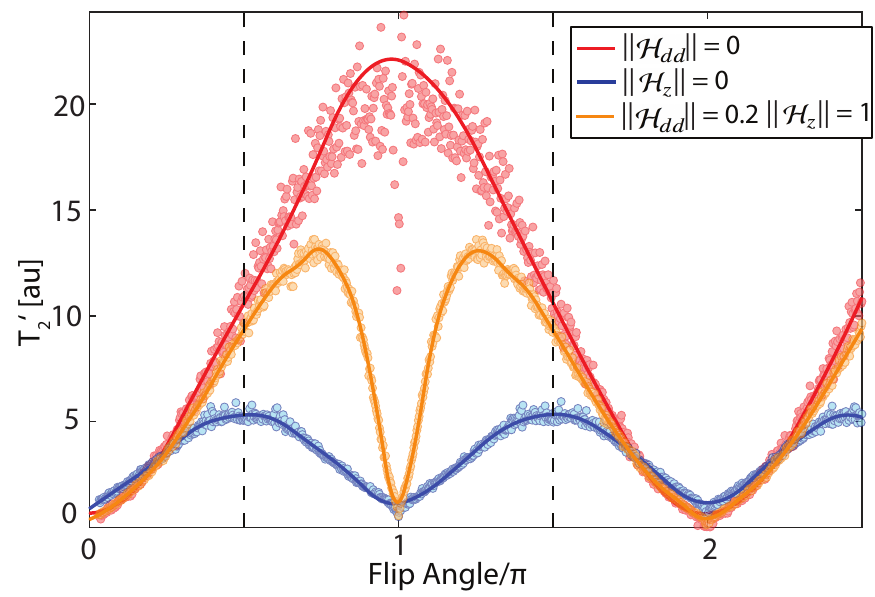}}
  \caption{\T{Intuitive picture of the $T_2'(\xt)$ profiles.} Red line considers the simulated $T_2'(\xt)$ profiles in the case of purely dephasing Hamiltonians ($\|H_{dd}\|=0$) applied to a $N_s=6$ spin network, resulting in an optimum at the CPMG condition ($\xt=\pi$). Similarly, the blue line considers case of pure dipolar interaction ($\|H_{z}\|=0$  giving a profile with optimum at the Waugh-Ostroff condition ($\xt=\pi/2$). Realistic spin networks considered in our experiments, for instance with $\|H_{dd}\|/\|H_{z}\|=0.2$ have a $T_2'(\xt)$ profile indicated by the yellow line. This can be considered as an interplay between the two limiting cases above, with a $\xtopt$ arising through the individual decay widths of the features at $\xt=\{\pi,2\pi\}$.}
\zfl{supp_WO}
\end{figure}

To simulate the dependence of $T_2'$ with flip angle $\xt$ of the applied pulses (displayed in \zfr{fig3}F of the main paper), we numerically simulate the action of the pulse sequence. For an accurate simulation, one has to consider the full many-body dynamics of the $>10^3$ $\Cs$ spins around every NV center that participate in this process. However to qualitatively gauge the system behavior, we consider full quantum simulations on relatively small networks consisting of $N_s=6$ spins. The $\Cs$ nuclei are arranged in a diamond lattice at concentrations corresponding to enrichment $\eta$ and we average the results over 50 random manifestations of the networks. We observe that even such small lattice sizes are able to qualitatively reproduce the features that we observe in the experiments (\zfr{fig3}F). 

We consider that the $\Cs$ spins are subject to both dephasing interactions $\mH_z$ as well as inter-spin dipolar coupling Hamiltonian $\mH_{dd}$. The relative dipolar couplings are calculated exactly employing the lattice positions of the $\Cs$ nuclei. Simultaneously, a random on-site field $\mH_z$ is added to the spins, and the ratio of the matrix norms $\|\mH_{dd}\|/\|\mH_z\|$ is allowed to be a tunable parameter. Since $\mH_z$ arises primarily from interactions with the P1 centers in the lattice, this serves as a proxy to account for varying P1 concentration in the samples of interest, and allows us to probe different regimes of interest between those dominated by disorder ($\|\mH_{dd}\|<\|\mH_z\|$), and by the inter-spin coupling ($\|\mH_{dd}\|>\|\mH_z\|$). The latter, for instance, is the operational regime in the samples with high $\Cs$ enrichment (e.g. in \zfr{fig4}C of the main paper).

 To simulate the system dynamics, we start with the initial state specified by the density matrix $\xr_I=I_x$, and consider evolution under Hamiltonian $\mH = \mH_{dd} + \mH_z$. We make the simplifying assumption that the pulses are applied on-resonance and are $\xd-$like, which is reasonable since $t_p\ll \|\mH\|^{-1}$ in our experiments. We calculate the final state $\xr_F = U^{\dg}\xr_IU$, where the propagator $U=[\exp(i\xt I_x)\exp(i\qt\mH)]^N$, where $N$ is the number of applied pulses. We assume $N$=2000, similar to the conditions employed in the experiments. To evaluate decay of the spin population in the rotating frame, we make the simplifying approximation that the survival probability of the spins is mono-exponential, $F(N\qt)=\Tr{\xr_I^{\dg}\xr_F}/\Tr{\xr_F^{\dg}\xr_F}$. Assuming $F\sim\exp(-N\qt/T_2')$ we evaluate the decay time constant $T_2'$ through the matrix logarithm, $T_2'\app-N\qt/\log(F)$. We note that this is only qualitatively valid; since the network sizes considered in the simulations are small and Hilbert space limited, the decay of transverse states may not be complete.

The simulation results are displayed in \zfr{fig3}F of the main paper, and in more detail in \zfr{supp_Hdd} and \zfr{supp_Hz}. The latter two panel sets unravel the relative contributions of dipolar interactions ($\|\mH_{dd}\|$) and on-site terms ($\|\mH_{z}\|$) to the decay. Here we artificially weight the two Hamiltonian norms $\|\mH_z\|$ and $\|\mH_{dd}\|$ with respect to each other, and focus on the disorder dominated regime, $\|\mH_{dd}\|<\|\mH_z\|$. We fit Gaussians to the features close to $\xt=\{\pi,2\pi\}$ (blue solid lines) as a guide to the eye, and to illustrate the dependence to the widths of the individual features.

We find that these simulations, even when carried out on relatively small network sizes, are able to qualitatively reveal identical features as in the experiments (see \zfr{fig3}A of main paper). From the panels in \zfr{supp_Hdd} where we consider the effect of increasing inter-spin dipolar coupling strength, the width of the decay feature near $\xt=\pi$ increases with $\|\mH_{dd}\|$, while the feature near $\xt=2\pi$ is only weakly affected. This points to origin of the $\xt=\pi$ feature from $\|\mH_{dd}\|$. Alternately, increasing $\mH_{z}$ predominately affects the width of the broad feature near $\xt=2\pi$ (see \zfr{supp_Hz}), while leaving the $\xt=\pi$ feature unaffected. While these panels consider the limit of weak-coupling, calibration of the weights of the terms in the alternate regime ($\|\mH_{dd}\|>\|\mH_z\|$) reveals additional dips around $\xt=\{2\pi/3,4\pi/3\}$. Using, $\|\mH_{dd}\|=4\|\mH_z\|$, for instance, we are qualitatively able to reproduce the experimental data for the higher enriched $\eta=$10\% sample in \zfr{fig4}F of the main paper. 

To intuitively understand the origin of the optimal flip angle $\xtopt$, we consider similar simulations in the limiting cases of pure dephasing and pure dipolar coupling (see \zfr{supp_WO}). In particular, the blue curve in \zfr{supp_WO} simulates the condition $\|\mH_{z}\|=0$ and displays $T_2'(\xt)$ profile with an optimum at $\xt=\pi/2$ (Waugh-Ostroff). The red curve in \zfr{supp_WO} simulates the condition $\|\mH_{dd}\|=0$ and yields a profile with an optimum at $\xt=\pi$ (CPMG). The yellow curve simulates the situation similar to the experimental condition, where $\|\mH_{z}\|=1,\|\mH_{dd}\|=0.2$, and shows that it arises as an interplay between the two limiting cases considered.

\section{Sequence construction}
\begin{figure}[t]
  \centering
  {\includegraphics[width=0.48\textwidth]{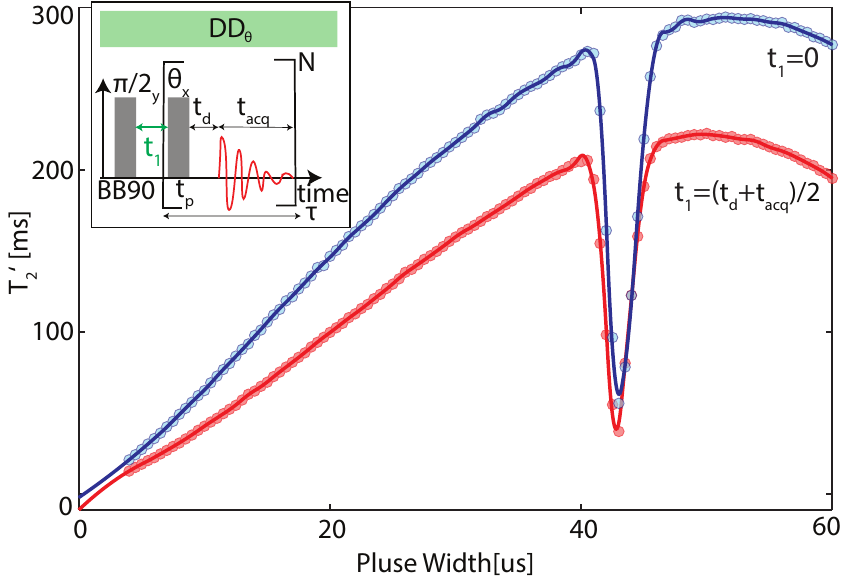}}
  \caption{\T{Experimental study of CPMG variants} under DD$_{\xt}$ sequence. We consider the effect when the first delay (see inset) $t_1=0$ (blue line) and $t_1=(\tacq+t_d)/2$ (red line), the latter corresponding to the exact CPMG condition when $\xt=\pi$. Both $T_2'(\xt)$ profiles have qualitatively the same shape, demonstrating that the theory developed is independent of the first delay. In these experiments, $t_p(\pi)=44\mu$s, $t_d=8\mu$s, and $\tacq=32\mu$s.}
\zfl{CPMG}
\end{figure}

We note that the sequence in Fig. 2A of the main paper (shown in inset of \zfr{CPMG}) can be considered as generally parameterizing commonly employed dynamical decoupling control sequences. If $t_p$ is the pulse width, $\tacq$ the acquisition time, $t_d$ the dead time, and $t_1$ the first pulse delay (see inset of \zfr{CPMG}), then spin locking entails ($\tacq=0, t_d = 0$), Waugh-Ostroff ($t_1=0$, $\xt=\pi/2$) and CPMG ($\xt=\pi$ and $t_1=(\tacq+t_d)/2$). In the experiments of Fig. 3-4 of the main paper, to maintain uniformity, we employ no delay between the first $\pi/2$ pulses and the applied control sequence, i.e. $t_1=0$. We still refer to the $\xt=\pi$ as being the CPMG condition, although it introduces minor modifications to filter-based description of CPMG dynamical decoupling. To confirm the notion that this does not affect the experimental results meaningfully, in \zfr{CPMG} we consider the $T_2'(\xt)$ profiles when $t_1=(\tacq+t_d)/2$ and $t_1=0$. The shapes of the obtained $T_2'(\xt)$ profiles remain qualitatively identical in both cases.

\section{Data Processing}
We detail here the acquisition and processing of the $\Cs$ NMR data under the applied DD$_{\xt}$ sequence. The inset in \zfr{fig2} of the main paper shows the acquired data in windows of period $\tacq$, with an interpulse interval $(t_p + t_d)$, where $t_d=6\mu$s is the dead time inserted to allow for ringdown of the probe and a delay for receiver and amplifier switching events. Total memory limitations (250k complex points) restrict the total decay acquisition periods, and for a suitable compromise, we use a relatively small number of points ($\sim$40) in every $\tacq$ window.

We make the approximation that $\tacq\ll \{T_2^{\ast},T_2'\}$, allowing us to average over the data in every $\tacq$ period. We use a median measure in this case to prevent against outliers due to acquisition data spikes that arise randomly on account of amplifier switching. The data, then separated by $\qt=(t_p + t_d+\tacq)$ are fit to extract decay time constants. We predominantly use stretched exponential fits; recent work has suggested that these are characteristic of slow dipolar decay~\cite{bordia2017probin}. We find similar behavior with biexponential fits with a definition of $T_2'$ as arising from the intersection with the $1/e$ value of the signal maximum. The graphs, while appearing qualitatively identical have slightly decreased $T_2'$ values in this case. To calculate the error bars in the $T_2'$ decay time constants in \zfr{fig3} and \zfr{fig4} of the main paper, we produce a spread of decay curves at 95\% confidence intervals given by the errors in the decay constant and the stretching factor. The error bars are then calculated by the spread upon intersection of a horizontal line at $1/e$ of the maximum value. Similarly, the error bars in the optimal flip angle $\xtopt$ value is evaluated from the data in \zfr{fig4} by determining the spread of the $\xt$ values that are within 95\% error bounds of the maximum measured $T_2'$ value. These values are tabulated in \ztr{table} of the main paper.

To calculate the SNR of the experimental data in \zfr{fig2}C, we make an assumption that the noise picked up by the NMR receiver is predominantly white in the frequency range being considered. The time domain data is Fourier transformed and SNR is defined as the ratio of the peak amplitude of the signal (at zero frequency) and the standard deviation of the spectrum 2kHz away from the peak (corresponding to over a thousand linewidths). We estimate then an SNR of $\app2\zt 10^4$ (\zfr{fig2}D). Note that memory limitations restrict the data collection to only 300ms in this case. With the ability to increase the total sampling memory and sampling rate, substantial improvement in SNR, by at least another two orders of magnitude beyond this value is possible.

  We note that in \zfr{fig2}D of the main paper, we process the regular FID in the same manner in order to make a comparison under the identical processing conditions. This provides the $\sim$533x SNR boost as indicated in \zfr{fig2}D. We note that since the total measurement times of the FID and DD$_{\xt}$ acquisition is identical here, the	ratio of SNR values is further boosted since after $T_2^{\ast}$, the normal FID measurement predominantly only picks up noise. Measurements that are optimized for SNR in both cases entail measurements only unto $T_2^{\ast}$ or $T_2'$ (respective decay constants), under which the SNR boost approaches $\lsb (1-\eta_d)T_2'/T_2^{\ast}\rsb^{1/2}\app$46. This constitutes the lower bound for the SNR boost under the applied sequence.

\begin{figure}[t]
  \centering
  {\includegraphics[width=0.49\textwidth]{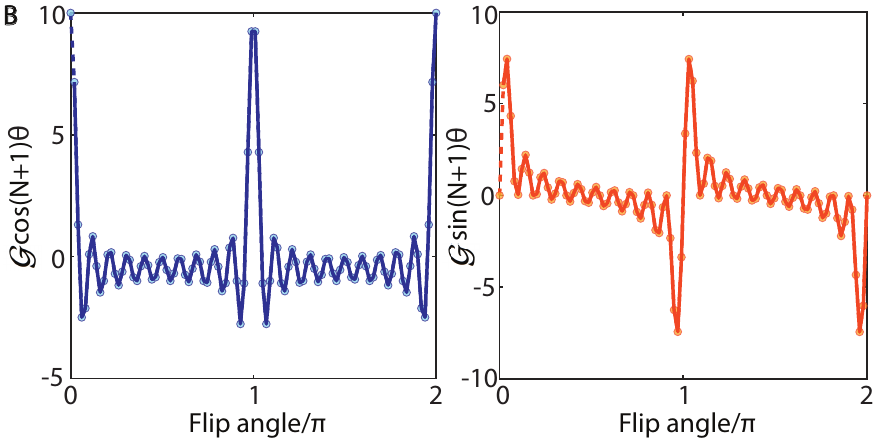}}
  \caption{\T{Time-domain diffraction grating} terms (A) $\mG(\xt)\cos(N+1)\xt$, and (B) $\mG(\xt)\sin(N+1)\xt$ in Eq. 1 of the main paper. Here we assume application of $N=$10 cycles of the control sequence. The filtering action in the zeroth order average Hamiltonian is evident for $\xt\neq n\pi$; both terms then show a fast decay that scales $\propto 1/N$ with increasing pulse number. On the other hand, at $\xt=n\pi$ the terms are unaffected. }
\zfl{grating}
\end{figure}

\section{Average Hamiltonian analysis}
In this section, we provide more details of the average Hamiltonian analysis of the DD$_{\xt}$ sequence, with an evaluation of zeroth and first order terms in the Magnus expansion. We note that we consider the limit of delta pulses, assuming negligible system evolution under the pulses. This is reasonable under the assumption $\|\mH\|t_p\ll1$, and is the operational regime in our experiments. We assume the pulses are applied exactly on-resonance, with a carrier frequency $\xo_0$. Going into the rotating frame with respect to the $\xo_0$, the net Hamiltonian of the $\Cs$ spins in the rotating frame is $\mH = \mH_z + \mH_{dd}$. For completeness, we specify here again the inter-nuclear dipolar coupling term $\mH_{dd} = \sum_{j<k} d_{jk}^{\CC}(3I_{jz}I_{kz} - \vec{I_j}\cdot\vec{I_k})$, where $I$ refers to spin-1/2 Pauli matrices, and the coupling strength,
$
d_{jk}^{\CC} = \fr{\mu_0}{4\pi}\hbar\xg_n^2(3\cos^2\xt_{jk}-1)\fr{1}{r_{jk}^3}
$
where, $\xg_n$=10.7MHz/T is the nuclear gyromagnetic ratio, and $\xt_{jk}=\cos^{-1}\lb \fr{\T{r}_{jk}\cdot\T{B}_0}{r_{jk}B_0}\rb$ is the angle of the interspin vector to the magnetic field. Note that $\|\mH_{dd}\|\sim$1kHz, and $t_p\app$50\us, and $\|\mH_{dd}\|t_p\ll1$.
Similarly, we have, $\mH_z = \sum_j c_jI_{jz}$, which arises on account of the interaction of the $\Cs$ nuclei with the electronic spin bath in the lattice.

In the rotating frame with respect to the $\xo_0$, evolution under the DD$_{\xt}$ sequence can be evaluated as the action of the unitary propagator on the $\Cs$ nuclei as, $U(N\qt) = [\exp(i\xt I_x)\exp(i\qt\mH)]^N$, where $N$ denotes the total number of pulses, and $\xt=\xO t_p$ is the pulse flip-angle under a given nuclear Rabi frequency $\xO$. Note that since we have assumed that the pulses are delta like, the effective evolution period is defined by the total pulse cycle length $\qt=(t_p+\tacq+t_d)$, where we include the pulse width $t_p$, free evolution time $\tacq$, and dead time $t_d$. Starting with an initial state $\xr_I\sim I_x$, the final state then is $\xr_F(N\qt)=U(N\qt)^{\dg}\xr_IU(N\qt)$, giving the final survival probability $F(N\qt)=\fr{1}{2}\Tr{\xr_F(N\qt)^{\dg}\xr_I}$.

Consider the propagator $U$, rewriting, we have,
\bea
U&=&\lsb e^{i\xt I_x}\exp(i\qt\mH)e^{-i\xt I_x}\rsb\lsb e^{i2\xt I_x}\exp(i\qt\mH)\right.\non\\
&\zt&\left.e^{-i2\xt I_x}\rsb\cdots e^{i(N+1)\xt I_x}\:,
\eea
that can be expressed in the form,
\beq
U = \prod_n \exp(i\qt\mH^{(n)}),
\eeq
where the toggling frame Hamiltonians, $\mH^{(n)}= e^{in\xt I_x}\mH e^{-in\xt I_x}$. One could evaluate the evolution as occurring from an effective average Hamiltonian, where the different orders are given by the Magnus expansion,
\beq
\ov{\mH} =\ov{\mH}^{(0)} + \ov{\mH}^{(1)} + \cdots
\eeq
with the zeroth and first order terms respectively,
\begin{eqnarray}
\ov{\mH}^{(0)} &=&  \frac{1}{N} \sum_{n=1}^{N} \mH^{(n)}\label{zeroth} \\
\ov{\mH}^{(1)} &=&  -\frac{i}{2N} \qt\sum_{n=1}^{N-1} \sum_{\ell=n+1}^{N}  [\mH^{(n)},\mH^{(\ell)}]  \zl{first}
\end{eqnarray} 
The Magnus expansion can be considered to converge if $\qt\|\mH\|\ll$1, which is the operational regime in our experiments. In the main paper, we had predominantly evaluated sequence performance under the zeroth order average Hamiltonian. Here we provide more detailed steps in the evaluation of these terms. Later, we also calculate the first order average Hamiltonian and comment on the validity of the zeroth order term over the range of flip angles considered. 

\T{\I{Dipolar interaction:}} -- Consider first the effective dipolar Hamiltonian arising from the action of the pulses on $\mH_{dd}$. The individual toggling frame Hamiltonians can in this case be evaluated as,
\bea
\mH_{dd}^{(n)} &=& \sum_{j<k}d_{jk}^{\CC}\lb 3\lsb I_{jz}I_{kz}\cos^2n\xt + I_{jy}I_{ky}\sin^2n\xt \right.\right.\non\\
&+&\left.\left. (I_{jz}I_{ky}+I_{jy}I_{kz})\cos n\xt\sin n\xt\rsb - \vec{I_j}\cdot\vec{I_k}\rb
\eea
Using the fact that $\sum_{k=1}^{N}\cos 2k\xt = \lb\fr{\sin N\xt}{\sin\xt}\rb\cos(N+1)\xt$ and $\sum_{k=1}^{N}\sin 2k\xt = \lb\fr{\sin N\xt}{\sin\xt}\rb\sin(N+1)\xt$, we can calculate the zeroth order average Hamiltonian as,
\begin{widetext}
\beq
\ov{\mH}^{(0)}_{dd} = \sum_{j<k}d_{jk}^{\CC}\lb \fr{3}{2}\lsb \mH_{\R{ff}} + \mH_{\R{dq}}\fr{1}{N}\lb\fr{\sin N\xt}{\sin\xt}\rb\cos(N+1)\xt + \wt{\mH}_{\R{ff}}\fr{1}{N}\lb\fr{\sin N\xt}{\sin\xt}\rb\sin(N+1)\xt\rsb -\vec{I_j}\cdot\vec{I_k}\rb
\zl{zero}
\eeq
\end{widetext}
where we define the flip-flop, double-quantum and tilted flip-flop Hamiltonians respectively as, 
\bea
\mH_{\R{ff,dq}} &=& I_{jz}I_{kz}\pm I_{jy}I_{ky}\non\\
\wt{\mH}_{\R{ff}}&=&I_{jz}I_{ky}+ I_{jy}I_{kz}
\eea
Let us define the grating function $\mG(\xt) = \fr{1}{N}\lb\fr{\sin N\xt}{\sin\xt}\rb$. The function resembles an optical diffraction grating, with $\mG(\xt)\rt 0$ for $\xt\neq n\pi$, peaks of $\mG(n\pi) = 1$, and a linewidth that falls $\sim 1/N$. 
The functional forms of these terms as they appear in \zr{zero} are shown in \zfr{grating} for $N$=10. The sharp filter-like dependence is evident. 
Hence, under the application of a large number of pulses, for instance $N>2000$ as in our experiments, the average Hamiltonian is transformed as,
\beq
\ov{\mH}^{(0)}_{dd}\app \sum_{j<k}d_{jk}^{\CC}\lb \fr{3}{2}\mH_{\R{ff}}-\vec{I_j}\cdot\vec{I_k}\rb.
\eeq
The zeroth order average Hamiltonian then evaluates to a simple flip-flop Hamiltonian that commutes with the initial state, $[I_x,\ov{\mH}_{dd}]=0$. This effectively locks the spins against decay in the rotating frame. This simple zeroth order average Hamiltonian already is able to capture the physics of the strong observed decays at the CPMG condition ($\xt=\pi$) and at $\xt=2\pi$. In both cases, the full dipolar coupling remains operational on the spins yielding rapid decay into many-body terms that are unobservable and manifest as decay. It also points to the fact that importantly, the CPMG condition is far from optimal for protecting the spins against decay while considering the presence of the interactions.

Let us now consider the first order average Hamiltonian. First defining
$\mH_{\R{Sff,Sdq}} = \sum_{j<k}d_{jk}^{\CC} \mH_{\R{ff,dq}}$ and
$\wt{\mH}_{\R{Sff}}= \sum_{j<k}d_{jk}^{\CC} \wt{\mH}_{\R{ff}}$,
we find the commutator between the toggling frame Hamiltonians, 
\bea
 &&[\mH_{dd}^{(n)},\mH_{dd}^{(\ell)}]  = \frac{9}{4} \lb(\cos (2\ell\xt) - \cos(2n\xt)) [\mH_{\R{Sff}},\mH_{\R{Sdq}}]\right. \non\\
&+&  \left.(\sin (2\ell\xt) - \sin(2n\xt))[\mH_{\R{Sff}},\wt{\mH}_{\R{Sff}}] +  \sin (2(\ell-n)\xt)  [\mH_{\R{Sdq}},\wt{\mH}_{\R{Sff}}]\rb \non
\eea
Using \zr{first}, we find the first order Hamiltonian
\bea
\ov{\mH}^{(1)}_{dd} &=& -\frac{9i}{4}\qt\left( f_1(\xt) [\mH_{\R{Sff}},\mH_{\R{Sdq}}] +   f_2(\xt) [\mH_{\R{Sff}},\wt{\mH}_{\R{Sff}}] \right.\non\\
&+&\left.  f_3(\xt) [\mH_{\R{Sdq}},\wt{\mH}_{\R{Sff}}] \right)\non 
\eea
where,
\begin{eqnarray}
f_1(\xt)  &=&  \frac{1}{2N}\sum_{n=1}^{N-1} \sum_{\ell=n+1}^{N}  ( \cos ( 2 l\xt) - \cos(2n\xt) ) \non \\
f_2(\xt)  &=&  \frac{1}{2N} \sum_{n=1}^{N-1} \sum_{\ell=n+1}^{N}  ( \sin ( 2 l\xt) - \sin(2n\xt)  )\non \\    
f_3(\xt)  &=&  \frac{1}{2N} \sum_{n=1}^{N-1} \sum_{\ell=n+1}^{N}   \sin ( 2 (l-n)\xt).  
\zl{filters}
\end{eqnarray} 

The modulating functions $f_1(\xt)$ , $f_2(\xt) $,  and $f_3(\xt)$ once again act as sharp filters, and have functional shapes that are easiest to discern when evaluated numerically. In the limit of  large $N$  the  functions vanish for all $\xt$ except near the points  $\xt = \{0, \pi, 2\pi\}$,  therefore $\ov{\mH}^{(1)}_{dd} \rightarrow 0$  for most flip angles.   At the points  $\xt = \{0, \pi, 2\pi\}$ exactly,  $f_1(\vartheta) = f_2(\vartheta) = f_3(\vartheta) = 0$  leading to a vanishing first order term,  $\ov{\mH}^{(1)}_{dd} = 0$, and giving rise to an exact expression, $\ov{\mH}^{(0)}_{dd} = \mH_{dd}$.

\T{\I{Dephasing spin-bath interaction:}} -- Let us now consider a similar average Hamiltonian analysis to zeroth and first order for the dephasing Hamiltonian arising out of interactions between the $\Cs$ nuclei and the spin bath of P1 centers. We have,
\begin{widetext}
\bea
\ov{\mH}^{(0)}_{z} &=& \sum_jc_{j}\lsb I_{jz}\fr{1}{N}\lb\fr{\sin N\xt/2}{\sin\xt/2}\rb\cos(N+1)\xt/2 +I_{jy}\fr{1}{N}\lb\fr{\sin N\xt/2}{\sin\xt/2}\rb\sin(N+1)\xt/2\rsb\non\\
&=& \sum_j\mG(\xt/2)c_{j}\lsb I_{jz}\cos(N+1)\xt/2 +I_{jy}\sin(N+1)\xt/2\rsb
\eea
\end{widetext}
Considering again the condition as before, when $\xt\neq 2n\pi$, $\labs\mG(\xt/2)\rabs\rt 0$, and the dephasing Hamiltonian is decoupled. Note that this is also the case when $\xt=\pi$, as expected for dynamical decoupling under the CPMG sequence. Therefore, the protocol allows the opportunity to separate contributions to the spin decay primarily driven by the inter-spin interactions when $\xt=(2n+1)\pi$, and including the effect of dephasing when $\xt=2n\pi$. The first order average Hamiltonian can now be calculated using,
\beq
\left[\mH^{(n)}_z,\mH^{(\ell)}_z\right]  =  2i\sum_j c_j^2\sin( (\ell-n)\xt) I_{jx}
\eeq
giving,
\beq
\ov{\mH}^{(1)}_{z} =  2 \qt \sum_j  c_j^2  f_3 (\xt/2)  I_{jx}
\eeq
where as before the function $f_3(\xt/2)$ constitutes a filter as described in \zr{filters}, and which for large N vanishes for all $\xt$ except for $\xt=\{0,2\pi\}$.

\end{document}